# TOWARDS ASSESSING CRITICAL INFRASTRUCTURES' CYBER-SECURITY CULTURE DURING COVID-19 CRISIS: A TAILOR-MADE SURVEY

Anna Georgiadou, Spiros Mouzakitis and Dimitrios Askounis

Decision Support Systems Laboratory, National Technical University of Athens, Iroon Polytechniou 9, 15780 Zografou, Greece

***ABSTRACT***

*This paper outlines the design and development of a survey targeting the cyber-security culture assessment of critical infrastructures during the COVID-19 crisis, when living routine was seriously disturbed and working reality fundamentally affected. Its foundations lie on a security culture framework consisted of 10 different security dimensions analysed into 52 domains examined under two different pillars: organizational and individual. In this paper, a detailed questionnaire building analysis is being presented while revealing the aims, goals and expected outcomes of each question. It concludes with the survey implementation and delivery plan following a number of pre-survey stages each serving a specific methodological purpose.*

## 1. INTRODUCTION

Coronavirus disease 2019, widely known as COVID-19, is an infectious dis-ease caused by severe acute respiratory syndrome coronavirus 2 (SARS-CoV-2) [1]. The disease was first detected in late 2019 in the city of Wuhan, the capital of China's Hubei province [2]. In March 2020, the World Health Organization (WHO) declared the COVID-19 outbreak a pandemic [3]. To-day, with more than 11 million confirmed cases in 188 countries and at least half a million casualties, the virus is continuing its spread across the world. While epidemiologists argue that the crisis is not even close to being over, it soon become apparent that "the COVID-19 pandemic is far more than a health crisis: it is affecting societies and economies at their core" [4].

Terms such as "Great Shutdown" and "Great Lockdown" [5, 6, 7] have been introduced to attribute the major global recession which arose as an eco-nomic consequence of the ongoing COVID-19 pandemic. The first noticeable sign of the coronavirus recession was the 2020 stock market crash on the 20th February. International Monetary Fund (IMF) in the April World Economic Outlook projected global growth in 2020 to fall to -3 percent. This is a downgrade of 6.3 percentage points from January 2020, making the "Great Lockdown" the worst recession since the Great Depression, and far worse than the Global Financial Crisis [7]. According to the International Labour Organization (ILO) Monitor, published on 7th April 2020, full or partial lockdown measures are affecting almost 2.7 billion workers, representing around 81% of the world's workforce [8].

Organizations from different business domains and operation areas across the globe try to survive this unprecedented financial crisis by investing their hopes, efforts and working reality on information technology and digitalization. Employees are being encouraged and facilitated on teleworking while most products and services become available over the web while, in many cases, transforming and adjusting to current rather demanding reality. However, these organizations come up against another COVID-19 side-effect not that apparent: the cyber-crime increase.

The increase in the population percentage connected to the World Wide Web and the expansion of time spent online, combined with the sense of confinement and the anxiety and fear generated from the lockdown, have formulated a prosperous ground for cyber-criminals to act. Coronavirus has rapidly reshaped the dark web businesses, as buyers and sellers jump on the opportunity to capitalize on global fears, as well as dramatic shifts in supply and demand. Phishing emails, social engineering attacks, malware, ransomware and spyware, medical related scums, investment opportunities frauds, are only a few examples of the cyber-crime incidents reported during the crisis period [9, 10].

INTERPOL's Cybercrime Threat Response team has detected a significant increase in the number of attempted ransomware attacks against key organizations and infrastructure engaged in the virus response. Cybercriminals are using ransomware to hold hospitals and medical services digitally hostage; preventing them from accessing vital files and systems until a ransom is paid [11].

Cyber-security agencies, organizations and experts worldwide have issued recommendations and proposed safeguard measures to assist individuals and corporations to defend against cyber-crime. While the virus is dominating in every aspect of our daily lives and human interaction is being substituted by digital transactions, cybersecurity gains the role it was deprived from during the last years. The question that remains unanswered, given the circumstances, is: What are the COVID-19 pandemic cyber-security culture side-effects on both individual and organizational level?

This paper presents the design and delivery plan of a survey aiming to evaluate the cyber-security culture during COVID-19 pandemic in the critical infrastructure domain. Section 2 presents background information regarding the importance of public cyber-security surveys conducted over the years emphasizing on the variety and originality of their findings. Building upon their approach, a detailed methodology is presented in Sections 3 & 4, in an effort to develop a brief, targeted and comprehensible survey for the assessment of the cybersecurity readiness of organizations during the crisis with emphasis on employees' feelings, thoughts, perspective, individuality. In Section 5, we sketch the survey next steps towards its conduction and fruitful completion. Finally, Section 6 concludes by underlying the importance of our survey reasoning while focusing on the challenging scientific opportunities that arise from it.

## 2. BACKGROUND

Over the last decades, cybersecurity surveys have been a powerful asset for information security academics and experts seeking to explore the constantly transforming technological reality. Their aim has been to reveal the contemporary trends on cybersecurity threats, organizations investment priorities, solutions for cloud security, threat management, application security, security training and certifications, and so many other topics.

Initially, they were narrowed down and addressed to certain participants depending on the nature and specific goal of each survey. A lighthouse representative of this kind was the Computer

Crime & Security Survey conducted by the Computer Security Institute (CSI) with the participation of the San Francisco Federal Bureau of Investigation's (FBI) Computer Intrusion Squad. This annual survey, during its 15 years of life (starting from 1995 and reaching up to 2010), was probably one of the longest-running continuous surveys in the information security field [12]. This far-reaching study provided unbiased information and analysis about targeted attacks, unauthorized access, incident response, organizational economic decisions regarding computer security and risk management approaches based on the answers provided by computer security practitioners in U.S. corporations, government agencies, financial institutions, medical institutions and universities.

Following their lead, numerous organizations of the public and private sector are seeking revealing findings that shall assist them in calibrating their operations and improving their overall existence in the business world via cybersecurity surveys. Healthcare Information and Management Systems Society (HIMSS) focusing on the health sector [13]; ARC Advisory Group targeting Industrial Control Systems (ICS) in critical infrastructures such as energy and water supply, as well as in process industries, including oil, gas and chemicals [14]; SANS exploring the challenges involved with design, operation and risk management of ICS, its cyber assets and communication protocols, and supporting operations [15]; Deloitte in conjunction with Wakefield Research interviewing C-level executives who oversee cybersecurity at companies [16]; these being only some of the countless examples available nowadays.

Current trend in the cybersecurity surveys appears to be broadening their horizon by becoming available and approachable to individuals over the internet [17, 18]. Since their goal is to reach out and attract more participants, thus achieving a greater data collection and, consequently, enforcing their results, tend to be shorter, more comprehensive to the majority of common people and apparently web-based.

Recognizing the unique value of this undisputable fruitful security evaluation methodology and rushing from the special working and living circumstances due to the COVID-19 pandemic, we identified the research opportuning to evaluate how this crisis has affected the cybersecurity culture of both individuals and organizations across the suffering globe. Security threats, frauds, breaches & perils have been brought to the light, recommendations have been given and precautions have been made [19, 20, 21]. What about the cybersecurity culture and its potential scars from this virus? Addressing this concern was our aim when designing, conducting and analysing the survey presented in this paper.

## 3. SECURITY CULTURE FRAMEWORK

During the last months, we have been conducting a thorough scientific research related to cyber-security tools, solutions and frameworks with a clear focus on the human factor. We have benchmarked the dominant reveals on the field, classified their possibilities and analysed their core security factors. Having identified their gaps and overlaps, common grounds and differentiations and thoroughly studied several academic principles regarding information security, including technical analyses, algorithmic frameworks, mathematical models, statistical computations, behavioural, organizational and criminological theories, we have created a foundation combining the elements that constitute the critical cyber-security culture elements [22]. The suggested cybersecurity culture framework is based on a domain agnostic security model combining the key factors affecting and formulating the cybersecurity culture of an organization. It consists of 10 different security dimensions analysed into 52 domains assessed by more than 500 controls examined under two different pillars: the organizational and the individual level. This hierarchical approach is being presented in Figure 1 while Figure 2 lists the model dimensions per level.

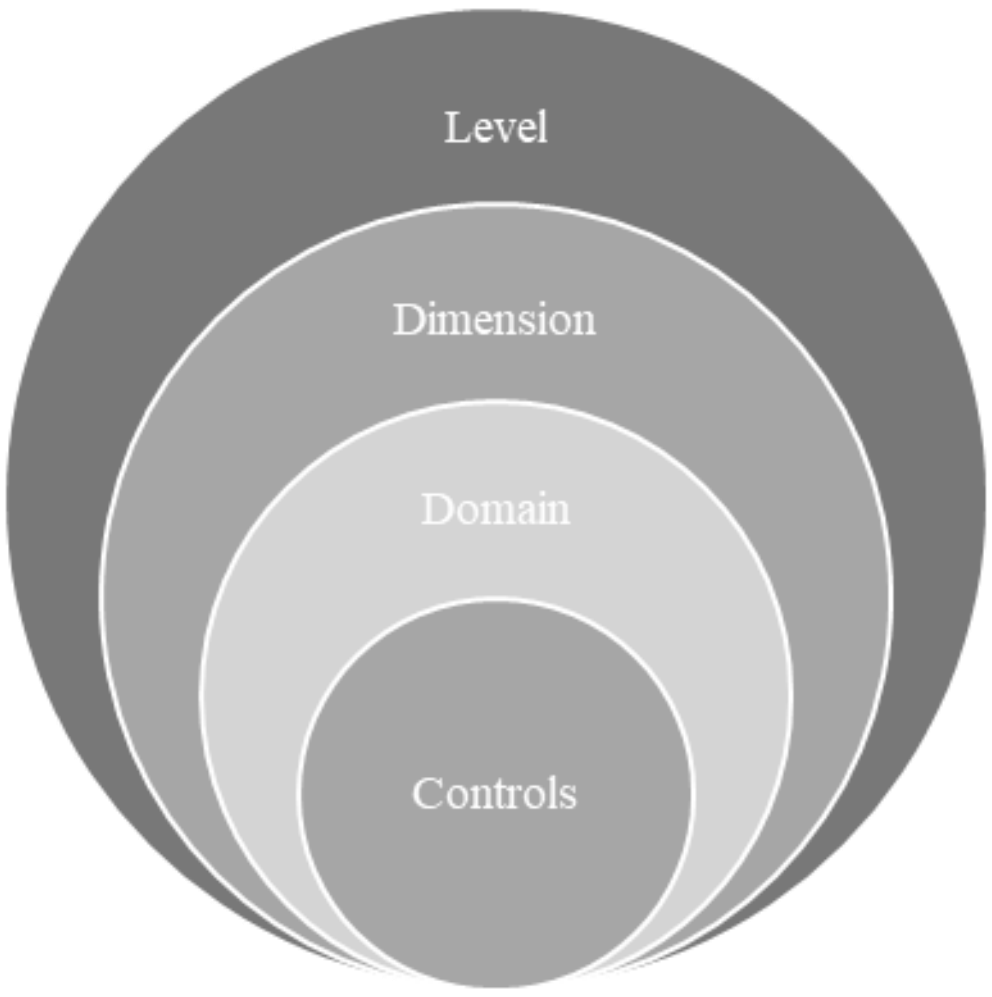

Figure 1. Cyber-Security Culture Model: Main Concepts

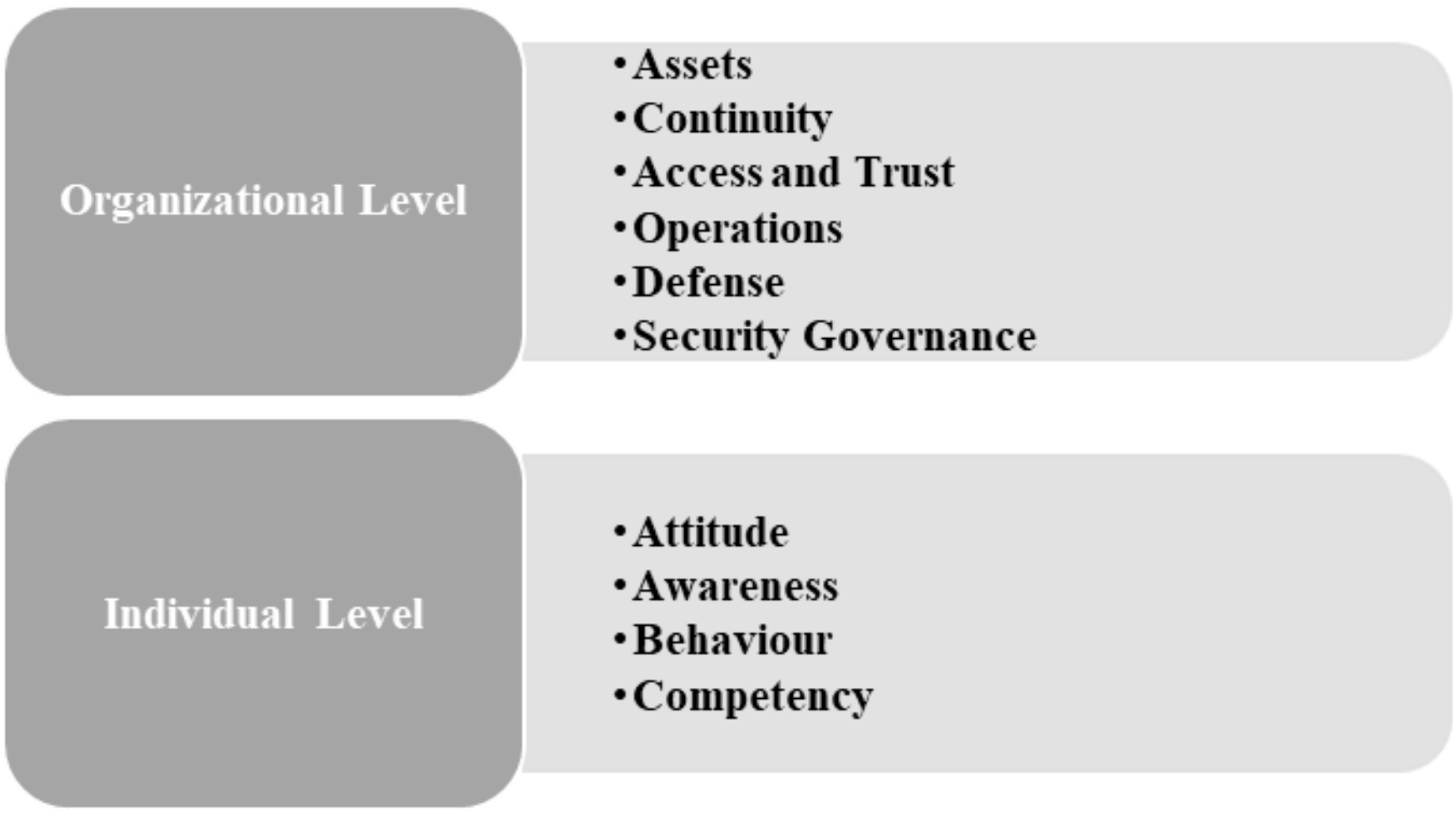

Figure 2. Cyber-Security Culture Model: Levels & Dimensions

An assessment methodology along with a weighting algorithm have been developed based on the previously mentioned model in order to offer a clockful cyber-security culture evaluation framework which shall be used as the basis of survey under design.

## 4. DESIGNING THE SURVEY

Our aim is to design a survey which shall be short and targeted getting the security pulse of current business reality in the critical infrastructure domain. One of our major goals is to keep the questionnaire small and easily addressed in a timely manner by a common employee with no special security expertise or knowledge. This way, we can facilitate participation of a broader workforce group minimizing effort and prerequisites while maximizing result variation and credibility. Towards that goal, we need to formulate questions to target specific security factors bridging various security domains while smartly extracting information depicting the existing working security routine and culture, their disruption by the COVID-19 crisis and their reaction to these special and rather demanding circumstances.

On the other hand, taking into consideration the reported cyber-crime incidents along with the fraud and attack techniques used by the criminals of the dark web during this period, we focused our evaluation on specific dimensions related to network infrastructure, asset management, business continuity, employee awareness and attitude.

In the paragraphs to follow, we outline how starting from a detailed cyber-security culture framework with more than 500 controls, we have narrowed down our objectives to a questionnaire containing no more than 23 questions, depending on the provided answers. Table 1 indexes the questions constituting the final version of our questionnaire including secondary clarification questions presented based on provided participant input whereas Table 2 correlates each of the questions to specific cyber-security levels, dimensions and domains of our model.

Table 1. Question indexing, including secondary clarification questions presented based on provided input (asterisk annotated).

| | | | | | |
|---|---|---|---|---|---|
| **Q1** | Prior to the COVID-19 crisis, were you able to work from home? | **Q9.2** | How were you informed how to use them? | **Q12.6** | I am proud to work for my organization. |
| **Q2.1** | Did you receive any security guidelines from your employer regarding working from home? | **Q10.1** | Has your company adopted a specific collaboration solution? | **Q12.7** | I have access to the things I need to do my job well. |
| **Q2.2*** | Please describe the main (2-3) security guidelines provided. | **Q10.2*** | What abilities does it offer? | **Q13** | What is your age? |
| **Q3** | What kind of devices are you using to connect to your corporate working environment? | **Q11.1** | Did you face any of the below cyber-security related threats during the COVID-19 crisis? | **Q14** | What is the highest degree or level of school you have completed? |
| **Q4** | Are these devices accessed by users other than yourself? | **Q11.2*** | Please name any other cyber-security threats you encountered during this period, not listed above. | **Q15** | Please select the business domain of the organization you work for. |
| **Q5** | These devices are personal or corporate assets? | **Q12.1** | To what extent do you agree with the following statements: (during this specific period of the COVID-19 crisis)<br><br>I prefer working from home than going to the office. | **Q16** | Which of the following best describes your work position? |
| **Q6** | Are these devices managed by your organization? | **Q12.2** | I work more productively from home. | **Q17** | Comments |
| **Q7** | Which of the following apply for the devices you currently use for your working from home employment? (providing security | **Q12.3** | I collaborate with my colleagues as effectively as when we are in office. | | |
| | measures alternatives, e.g. antivirus, password protections) | | | | |
| **Q8** | How do you obtain access to your corporate working environment? | **Q12.4** | I am satisfied by my employer's approach to the crisis. | | |
| **Q9.1** | Were you asked to use applications or services that you were unfamiliar with, because of the need for remote working? | **Q12.5** | I have all the support I need to face any technical problems I have (e.g. corporate access issues, infrastructure failures, etc.). | | |

### 4.1. Organizational Level

Culture is defined as a set of shared attitudes, values, goals and practices that define an institution or organization. Consequently, cyber-security culture refers to the set of values, conventions, practices, knowledge, beliefs and behaviours associated with information security. Therefore, its skeleton is being outlined by the working environment along with the technological infrastructure and security countermeasures that define it.

To understand, evaluate and analyse the security cultural status of the critical infrastructure organizations participating to our survey, we have included questions Q1-Q10 that heartbeat the overall technological and security readiness and adaptability. Under the coronavirus prism, we intend to understand if teleworking was possible prior to the crisis or not and under which security policies. Thus, we have included queries polling the remote access procedures and their meeting standards as well as the types, configuration and management of the devices used to gain access to the corporate environments. In other words, we attempt to assess the way and the means of the working from home experience with a clear focus on cyber-security.

Additionally, we intend to assess the security maturity of the management, the security response team and awareness training program by introducing a number of questions clearly related to cyber-security familiarity and readiness. The most critical question of these category is the one referring to security guidelines provided during the COVID-19 crisis seeking to match their responsiveness and contemporality. Informing your workforce by issuing supportive guiding principles following the example of leading cyber-security entities and experts during rather confusing and challenging periods is a core security indicator.

Another business facet which is examined, although not directly related to information security, is the collaboration possibilities offered to employees. Communication and teamwork need to be facilitated and promoted, especially during this time period when general isolation is mandated as the only defence against the virus spread. Companies are expected to provide all means necessary to assist their employees in being productive, effective and cooperative. This notion and quality are being tested via two simplified questions included into our survey.

### 4.2. Individual Level

Moving down to an individual level, evaluation becomes more demanding since virus fear and emotional stress dominate every aspect of daily life directly or indirectly affecting the human related security factors. Questions Q11-Q12 attempt to probe the security behaviour, attitude and competency of the remote workers by examining their emotions, thoughts and beliefs and by asking them to report any security incidents they came up against.

Questions Q13-Q16 refer to generic information used for an individual profiling and categorization which shall enable us to analyse gathered results under different prisms offering various grouping possibilities and leading to possibly interesting findings on the basis of age, industry, education, working experience and expertise.

Table 1. Correlating questions to cyber-security culture framework

| | **Organizational Level** | | | | | | **Individual Level** | | | |
|---|---|---|---|---|---|---|---|---|---|---|
| | **Assets** | **Continuity** | **Access and Trust** | **Operations** | **Defense** | **Security Governance** | **Attitude** | **Awareness** | **Behaviour** | **Competency** |
| **Q1** | - Network Infrastructure Management<br>- Network Configuration Management | | - Access Management<br>- External Environment Connections | | | | | | | |
| **Q2.1** | | Change Management | | Organizational Culture and Top Management Support | Security Awareness and Training Program | Security Management Maturity | | | | |
| **Q2.2*** | | | | | | | | | | |
| **Q3** | Hardware Assets Management | | Access Management | | | | | | | |
| **Q4** | | | Access Management | | | | | | - Policies and Procedures Compliance<br>- Security Behaviour | |
| **Q5** | - Hardware Assets Management<br>- Information Resources Management<br>- Data Security and Privacy | | - Access Management<br>- External Environment Connections | | | | | | | |
| **Q6** | - Hardware Assets Management<br>- Software Assets Management<br>- Information Resources Management<br>- Data Security and Privacy | | - Access Management<br>- External Environment Connections | | | | | | | |
| **Q7** | - Hardware Configuration Management<br>- Information Resources Management<br>- Data Security and Privacy | | | | Malware Defense | | | Policies and Procedures Awareness | - Policies and Procedures Compliance<br>- Security Behaviour<br>- Security Agent Persona | |
| **Q8** | - Network Infrastructure Management<br>- Network Configuration Management | | - Access Management<br>- External Environment Connections | | Boundary Defense | | | | | |
| **Q9.1** | | - Business Continuity & Disaster Recovery<br>- Change Management | | | | | | | | |
| **Q9.2** | | | Communication | Organizational Culture and Top Management Support | | | | | | |
| **Q10.1**<br>**Q10.2*** | | | | Operating Procedures | | | | | | |
| **Q11.1**<br>**Q11.2*** | | | | | | | | | - Security Behaviour<br>- Security Agent Persona | Security Skills Evaluation |
| **Q12.1**<br>**Q12.2**<br>**Q12.3** | | | | | | | Employee Climate | | | |
| **Q12.4**<br>**Q12.5**<br>**Q12.6**<br>**Q12.7** | | | | | | | Employee Satisfaction | | | |
| **Q13**<br>**Q14**<br>**Q15**<br>**Q16** | | | | | | | Employee Profiling | | | |

The accruing questionnaire manages to effectively and efficiently combine the two security levels of our framework. Additionally, its contents have been tailored to rapidly yet effectually heartbeat the cyber-security reality during a disrupting chronological period, such as the COVID-19 pandemic. This agile instrument, although offering a quick and fruitful measurement method compared to similar concurrent surveys, it cannot be considered an in-depth information security assessment. Furthermore, it should not be used to label participating organisations but only to provide an overview of current status.

## 5. NEXT STEPS

Having developed a first questionnaire version addressing the security elements of interest based on our security culture framework, we need to carefully design the rest of the survey methodology including:

- **validity testing**: identify ambiguous questions or wording, unclear instructions, or other problems prior to widespread dissemination possibly conducted by a group of survey experts, experienced researchers and analysts, certified security and technology officers.

- **delivery method**: select the appropriate delivery method and possibly run an instrument validity testing to verify survey conduction methodology

- **sample selection**: carefully chose representatives from energy, transport, water, banking, financial market, healthcare and digital infra-structure from different European countries (e.g. Cyprus, France, Ger-many, Greece, Italy, Romania, Spain) affected by the COVID-19 crisis.

- **survey duration**: defining a specific start and end period communicated to all invited parties.

## 6. CONCLUSIONS AND FUTURE WORK

Our survey focuses on evaluating the security readiness and responsiveness of corporations during the Great Shutdown and more specifically it shall be addressing critical infrastructure domain representatives from different countries affected by the coronavirus.

Security cultural approach demands flexibility and concurrency. In a radically evolving and transforming environment, security and risk teams need to become part of the crisis management group, remote working employees need to remain vigilant to cyber-threats and operations life-cycle needs to remain uninterrupted especially for the operators of essentials services. Our research aims to investigate if and in what extend is this approach embraced by the critical infrastructure organizations in different countries nowadays while revealing interesting findings related to cyber-security and inspiring further scientific researches on this field.

## ACKNOWLEDGEMENT

This work has been co-funded from the European Union's Horizon 2020 research and innovation programme under the EnergyShield project "Integrated Cybersecurity Solution for the Vulnerability Assessment, Monitoring and Protection of Critical Energy Infrastructures", grant agreement No 832907 [23].

## AUTHORS

**Mrs. Anna Georgiadou** is a research associate in the Management and Decision Support Systems Laboratory in the School of Electrical and Computer Engineering at the National Technical University of Athens (NTUA). She has been working as a senior engineer on operation and system support services for major telecommunication providers, energy regulators, banking systems and other critical national infrastructures. She has recently become a PhD candidate on the cyber-security field inspired by her active participation in the HEDNO's (Hellenic Electricity Distribution Network Operator) information security management group. She is a certified database and network administrator, software developer and data-analyst.

**Dr. Spiros Mouzakitis** is a senior research analyst for National Technical University of Athens (NTUA). He has 18 years of industry experience in conception, analysis and implementation of information technology systems. His research is focused on decision analysis in the field of decision support systems, risk management, Big Data Analytics, as well as optimization systems and algorithms, and enterprise interoperability.He has published in numerous journals including Computer Standards & Interfaces, International Journal of Electronic Commerce, Information Systems Management, and Lecture Notes in Computer Science, and has presented his research at international conferences.

**Dr. Dimitris Askounis** is a Professor in the School of Electrical and Computer Engineering at the National Technical University of Athens (NTUA). He has been involved in numerous IT research and innovation projects funded by the EU since 1988 in the thematic areas of eBusiness interoperability, eGovernance, data exploitation and management, decision support, knowledge and quality management, computer integrated manufacturing, enterprise resource planning, etc. He teaches digital innovation management, decision support and management systems, and he is a member of scientific committees on innovation and entrepreneurship competitions and incubators offered by International University networks, Financial Institutions, etc. Dr. Askounis has published over 80 papers in scientific journals and international conference proceedings.